\documentclass[english,aip,pof,preprint]{revtex4}
\usepackage[latin9]{inputenc}
\usepackage{float}
\usepackage{bm}
\usepackage{amsmath}
\usepackage{amssymb}
\usepackage{graphicx}

\makeatletter

\newcommand{\lyxdot}{.}

\@ifundefined{textcolor}{}
{%
 \definecolor{BLACK}{gray}{0}
 \definecolor{WHITE}{gray}{1}
 \definecolor{RED}{rgb}{1,0,0}
 \definecolor{GREEN}{rgb}{0,1,0}
 \definecolor{BLUE}{rgb}{0,0,1}
 \definecolor{CYAN}{cmyk}{1,0,0,0}
 \definecolor{MAGENTA}{cmyk}{0,1,0,0}
 \definecolor{YELLOW}{cmyk}{0,0,1,0}
}

\usepackage{babel}

\makeatother

\usepackage{babel}
\begin{document}

\title{Attracting fixed points for heavy particles in the vicinity of a
vortex pair}

\author{S. Ravichandran}

\email{ravis@tifrh.res.in}

\affiliation{TIFR Centre for Interdisciplinary Sciences \\
 Tata Institute of Fundamental Research, Narsingi, Hyderabad, 500075,
India.}

\author{Prasad Perlekar}

\email{perlekar@tifrh.res.in}

\affiliation{TIFR Centre for Interdisciplinary Sciences \\
 Tata Institute of Fundamental Research, Narsingi, Hyderabad, 500075,
India.}

\author{Rama Govindarajan}

\email{rama@tifrh.res.in}

\affiliation{TIFR Centre for Interdisciplinary Sciences \\
 Tata Institute of Fundamental Research, Narsingi, Hyderabad, 500075,
India.}
\begin{abstract}
We study the behaviour of heavy inertial particles in the flow field
of two like-signed vortices. In a frame co-rotating with the two vortices,
we find that stable fixed points exist for these heavy inertial particles;
these stable frame-fixed points exist only for particle Stokes number
$St<St_{cr}$. We estimate $St_{cr}$ and compare this with direct
numerical simulations, and find that the addition of viscosity increases
the $St_{cr}$ slightly. We also find that the fixed points become
more stable with increasing $St$ until they abruptly disappear at
$St=St_{cr}$. These frame-fixed points are between fixed points and
limit cycles in character.
\end{abstract}
\maketitle

\section{Introduction\label{sec:Introduction}}

Vortices are building blocks of turbulent fluid flow. During their
evolution, vortices stretch, rotate and interact with other vortices.
Energy in turbulent flows is transferred to larger and smaller scales
by vortex mergers and stretching respectively. In the Earth's atmosphere,
in industrial processes and in water bodies, turbulent flows often
carry particles such as dust or aerosols with them. Such particles
are typically heavier than the fluid which carries them, and this
paper is devoted to the effect of vortices on heavy particles. A large
number of simulations and experiments \cite{Davilla2001,Benczik2002,Bec2003,Chen2006,Derevyanko2007,Goto2008,Tallapragada2008,Toschi2009,Eidelman2010,Perlekar2011,Gibert2012}
have shown that the transport of inertial particles in two- and three-dimensional
turbulence is not like the transport of inertialess particles. In
particular, inertial particles cluster. The primary reason for this
in a turbulent flow is vorticity, and the tendency of heavy inertial
particles to leave the neighborhood of a vortex. What happens when
there is more than one vortex? Could heavy particles in fact have
long residence times near vortices?

We choose the simplest flow with more than one vortex, namely the
flow generated by two identical vortices of the same sign at a distance
$2R$ from each other. Kinematics dictates that these vortices, when
there is no viscosity, will cause each other to rotate around the
origin at an angular velocity $\Omega$. This flow, for particles
of a given small Stokes number, is shown below to display `fixed'
points in a moving frame of reference. An important consequence of
this is that particles cluster in a location of low vorticity, but
close to the vortices. These cluster points rotate with the flow.
The location of cluster is different for different Stokes number,
and vanishes beyond a critical Stokes number. The behavior is first
analyzed in an inviscid framework, and followed by viscous simulations
to show that the same behavior is displayed there as well.

The rest of the paper is organized as follows. We first describe our
system and approach, and then present an analytical investigation
of heavy particles in the like-signed vortex-pair flow. This is followed
by a presentation of our point vortex simulations and a linear analysis.
We end with a discussion of our direct numerical simulations, and
conclusions.

\section{Elliptic Fixed Points in Lab and Rotating Frames of Reference}

We begin by describing the system and the equations we use for heavy
particles -- first in the lab-frame and then in a frame rotating with
the vortices -- and then discuss why elliptic fixed points in a lab-fixed
frame of reference are very different from those in a rotating frame
of reference. Heavy particles cannot cluster in the vicinity of the
former as proved in Ref.\cite{Sapsis2010} but we show that they can
and do cluster in the vicinity of the latter.

In the lab frame, we have two like-signed point vortices, each of
circulation $\Gamma$ and placed at a distance of $2R$ from the other,
describing motion on a circle of radius $R$, with a time period $T=8\pi^{2}R^{2}/\Gamma$.
The vortices remain in antiphase from each other. The fluid velocity
at a point $\mathbf{r}$ is given by: 
\begin{eqnarray}
\mathbf{u}_{\mbox{lab}} & = & \frac{d\mathbf{r}}{dt}=\frac{\Gamma}{2\pi}\mathbf{e}_{z}\times\left(\frac{\mathbf{r}-\mathbf{R}}{\left|\mathbf{r}-\mathbf{R}\right|^{2}}+\frac{\mathbf{r}+\mathbf{R}}{\left|\mathbf{r}+\mathbf{R}\right|^{2}}\right)\mbox{,}\label{velocity field, lab frame}
\end{eqnarray}
 where the vortices are at $\pm\mathbf{R=}\left(\pm X,\pm Y\right)$,
and $\mathbf{e}_{z}$ is the unit vector pointing out of the page.
The subscript `${\mbox{lab}}$' denotes that something is measured
in the lab-fixed frame of reference. 

The motion of rigid, inertial point particles is modeled by using
the Maxey-Riley equations \cite{Maxey1983} in the lab frame: 
\begin{eqnarray}
\frac{d\mathbf{\mathbf{x}}}{dt} & = & \mathbf{\mathbf{v}}\nonumber \\
\frac{d\mathbf{\mathbf{v}}}{dt} & = & -\frac{1}{St}(\mathbf{v}-\mathbf{u})+\beta\frac{D\mathbf{u}}{Dt}\label{eq:particle motion}
\end{eqnarray}
 where $\beta=3\rho_{f}/(2\rho_{p}+\rho_{f})$, $\rho_{p}$ and $\rho_{f}$
are the particle and fluid densities respectively, $\mathbf{v}$ and
$\mathbf{u}$ are the particle and fluid velocities respectively,
and $\mathbf{x}$ the location of a particle, $St=\tau/T$ is the
Stokes number, $\tau=\frac{2}{9}\frac{a^{2}}{\nu}\frac{\rho_{p}}{\rho_{f}}$
is the relaxation time of the particle, and $T$ is a characteristic
time scale of the flow, taken here to be the time period of rotation.
Light particles ($\rho_{p}\ll\rho_{f}$ i.e., $\beta\approx3$) cluster
in the regions of vortices whereas, heavy particles, such as aerosols,
$\rho_{p}\gg\rho_{f}$ ($\beta=0$) are expelled from vortical regions
\cite{Sapsis2010}.

\begin{figure}
\noindent \includegraphics[width=0.5\textwidth]{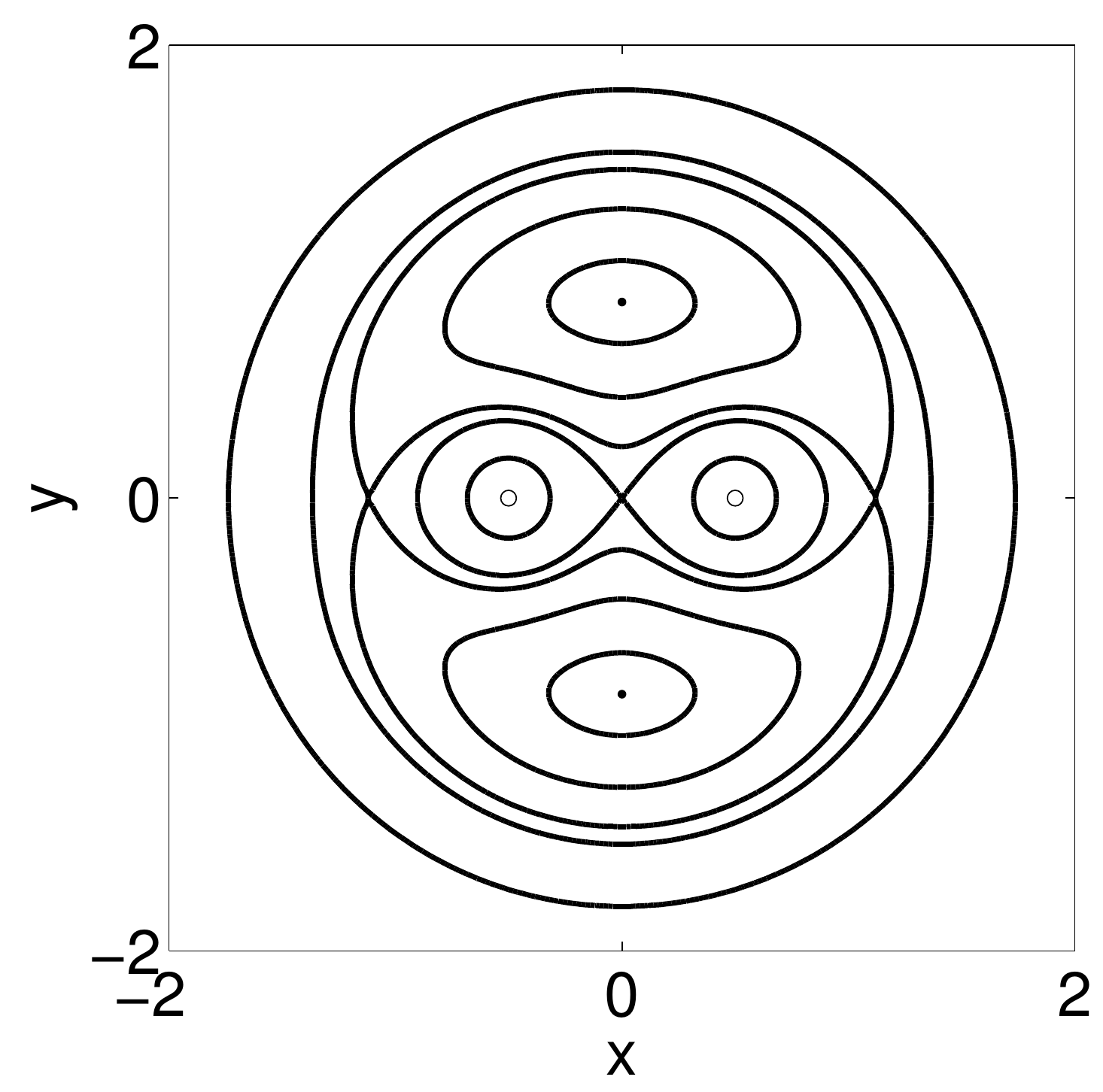}
\caption{Invariant manifolds in the rotating frame in the flow around two identical
point vortices. The vortices are indicated by the small circles at
$\left(\pm0.5,0\right)$, and the elliptic fixed points of our interest
by the filled circles at $\left(0,\pm\sqrt{3}/2\right)$.\label{fig:invariant manifolds}}
\end{figure}

In the frame of reference rotating with the same angular velocity
as the vortices {[}$\Omega=\Gamma/(4\pi R^{2})${]}, the flow field
is divided into water-tight compartments by the separatices -- invariant
manifolds, see e.g. \cite{Rom-Kedar1990,Newton2001} -- shown in figure
\ref{fig:invariant manifolds}. The equations of motion \eqref{eq:particle motion}
may be transformed to this rotating frame and are given below. Quantities
with a $\hat{}$ are measured in the rotating frame.

\begin{align}
\frac{d\hat{\mathbf{x}}}{d\hat{t}} & =\hat{\mathbf{v}}\nonumber \\
\frac{d\hat{\mathbf{v}}}{d\hat{t}} & =\frac{\hat{\mathbf{u}}-\hat{\mathbf{v}}}{St}-2\bm{\Omega}\times\mathbf{\hat{v}}+\Omega^{2}\mathbf{\hat{r}}\label{eq:Particle_motion_rotating_frame}
\end{align}

For $St=0$, the system has two fixed points (centers) at $(0,\pm\sqrt{3}R)$
in the rotating frame of reference. The elliptic nature of these fixed
points may be seen by linearizing the velocity field at these points
{[}see section \eqref{sub:Linear-analysis}{]}. These `fixed-points'
are actually rotating at the same angular velocity as the vortices
in the lab-fixed frame, so what we have are moving fixed points!

Incidentally we would like to distinguish our rotating frame of reference
from the one that is normally used to simulate, for instance, Earth's
rotation. In the latter, the rotation of the lab is imposed as an
additional forcing. Our lab is not rotating, but our flow is steady
in a non-inertial frame that rotates with the angular velocity of
the vortex pair. We also note that the points at $(0,\pm\sqrt{3}R)$
are fixed points for fluid particles and tracer particles, and not
necessarily for inertial particles.

As previously mentioned, elliptic fixed points in the rotating frame
of reference are qualitatively different from elliptic fixed points
in a lab fixed frame of reference. Sapsis \& Haller \cite{Sapsis2010}
show that the fixed points in a region of the flow consisting of elliptic
streamlines cannot be attractors for heavy particles of small Stokes
number. Their argument is that since $\beta\simeq0$ for heavy particles,
at $O\left(St\right)$ $\mathbf{u}$ may be used to replace $\mathbf{v}$
on the left hand side of eq. (\ref{eq:particle motion}), to get 
\begin{equation}
\mathbf{v}=\mathbf{u}-St\frac{D\mathbf{u}}{Dt}+\mathcal{O}\left(St^{2}\right),\label{eq: inertial equation}
\end{equation}
where $\frac{d\mathbf{v}}{dt}$ was replaced with $\frac{D\mathbf{u}}{Dt}$
in the second term. This is called the `inertial equation'. The divergence
of $\mathbf{v}$, since $\mathbf{u}$ is divergence-free in incompressible
flow, becomes 
\begin{equation}
\nabla\cdot\mathbf{v}=-St\nabla\cdot\left(\mathbf{u}\cdot\nabla\mathbf{u}\right)=-St\frac{\partial u_{i}}{\partial x_{j}}\frac{\partial u_{j}}{\partial x_{i}}=StQ\mbox{,}\label{eq:Okubo-Weiss}
\end{equation}
 where $Q=\left(\omega^{2}-s^{2}\right)$ is the Okubo-Weiss parameter,
$s$ and $\omega$ being the symmetric and anti-symmetric parts of
the strain-rate tensor respectively. A word of caution: in a general
turbulent flow we must remember that particle velocities do not form
a field, in the sense that there can be two very different particle
velocities at the same location, so strictly speaking we may not define
a quantity called divergence for particle velocity. However, at small
Stokes numbers, the inertial equation allows $\mathbf{v}$ to be approximated
to a field, and thus lets us relate this quantity to particle clustering
for negative divergence and to particles leaving the neighborhood
at positive divergence. At an elliptic fixed point in a lab-fixed
frame, we must have $Q>0$, and Liouville's theorem may be applied
to a small volume in its vicinity to show that it cannot attract heavy
particles. There is thus no attracting elliptic fixed point for small
Stokes number heavy particles in a fixed frame of reference.

\begin{figure}[!t]
\noindent \includegraphics[width=0.5\linewidth]{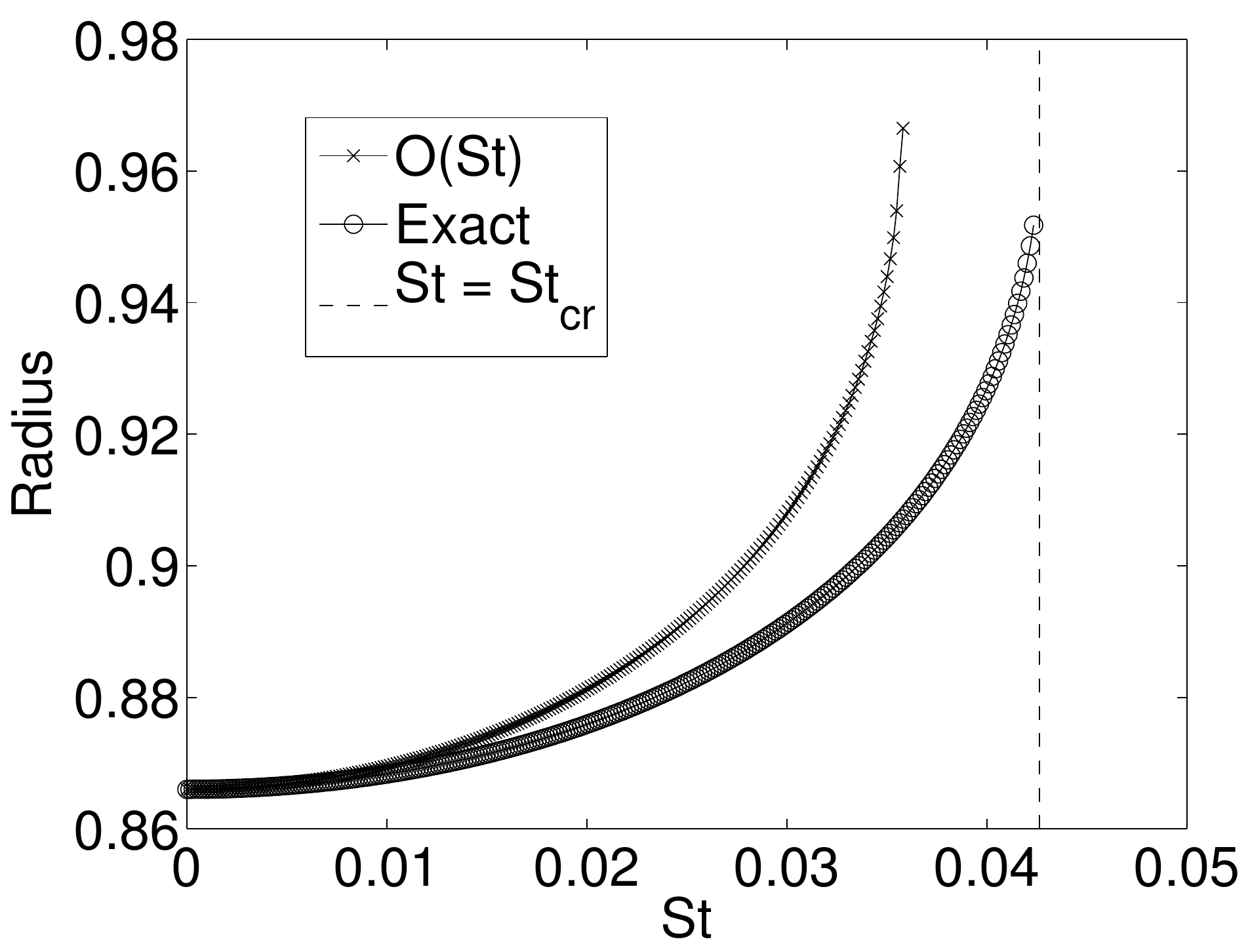}\includegraphics[width=0.5\linewidth]{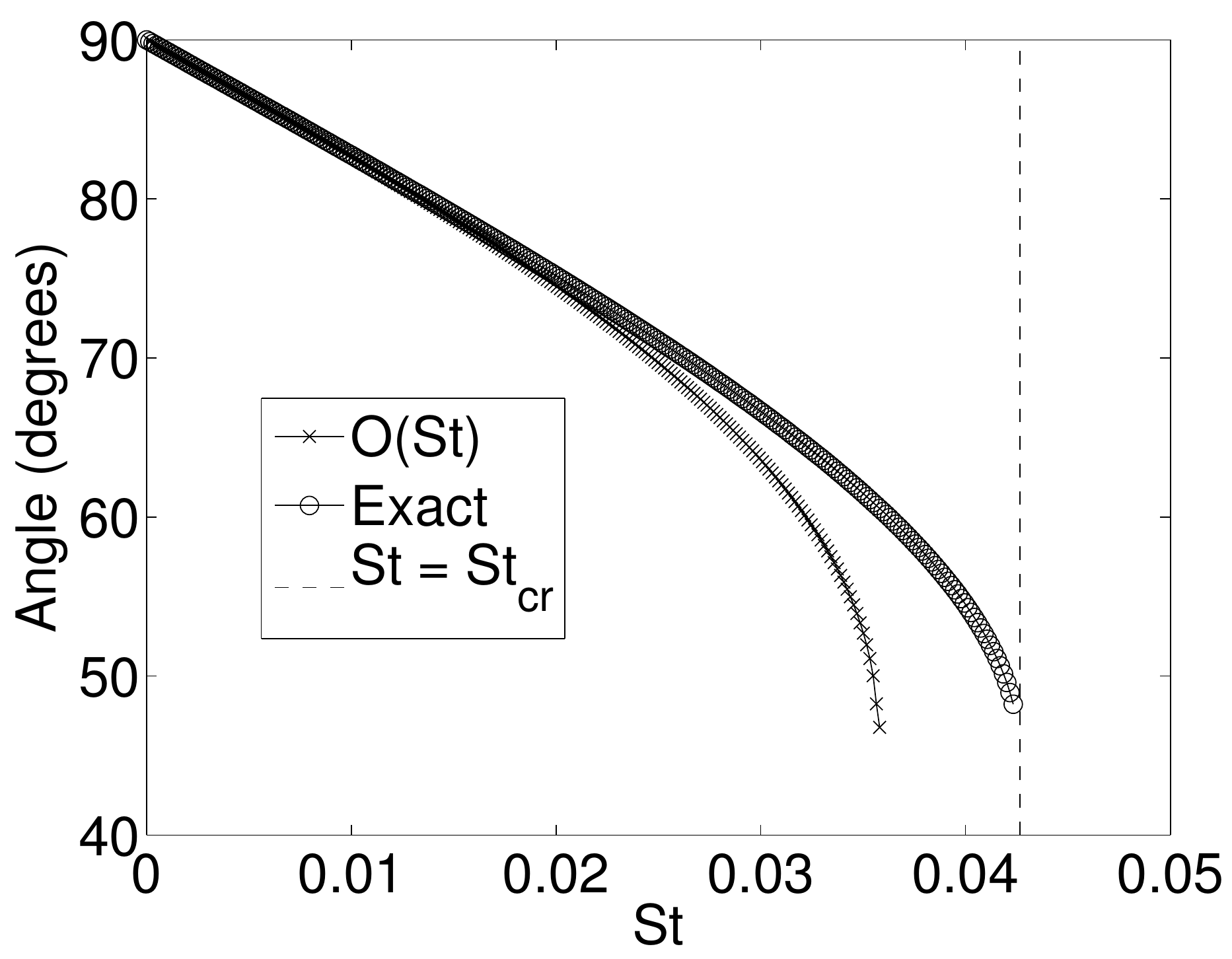}
\caption{\label{fig:Fixed-point-comparison}Locations of fixed points (in radians)
for inertial particles with different Stokes number obtained from
eq. \eqref{eq:fixed points exact} (circles). We also show the location
of fixed points obtained by using the small Stokes number ($\mathcal{O}(St)$)
approximation (cross). There are no fixed points for $St>St_{cr}$.
{[}Here and in all figures to follow, simulations were done with $\Gamma=2\pi$.
{]}}
\end{figure}

How about in a rotating frame of reference? We saw in the flow under
consideration that there are two elliptic fixed points for tracer
particles in the rotating frame. We next ask whether there are fixed
points for inertial particles as well, i.e., are there locations where
an inertial particle would remain forever. This may be done by solving
for fixed points of eq. \ref{eq:Particle_motion_rotating_frame},
giving 
\begin{equation}
\frac{\hat{\mathbf{u}}}{St}=-\Omega^{2}\mathbf{r}.\label{eq:fixed points exact}
\end{equation}
Equation (\ref{eq:fixed points exact}) has to be solved numerically.
This was done using the MATLAB function minimisation routine ``fsolve''.
Solutions for eq. \ref{eq:fixed points exact} exist for $St<St_{cr}=0.04264543$.
For $St>St_{cr}$, the minima of $\hat{\mathbf{u}}+St\Omega^{2}\hat{\mathbf{r}}$
are greater than zero; i.e., there exist no solutions. Solutions of
eq. \ref{eq:fixed points exact} lying in the top half-plane is shown
in figure \ref{fig:Fixed-point-comparison}. A symmetrically placed
fixed point exists with $x<0,y<0$. We see that the fixed points start
on the y-axis at $\sqrt{3}R$ for particles of $St=0$, and move progressively
away from the vertical as $St$ increases. Thus particles of increasing
inertia display fixed points which drift further and further away
from the axis of symmetry. One may imagine that with inertia, the
particles find it harder to ``keep up'' with the rotating frame
and drift in the opposite direction. Note that the fixed points lie
within the region of the flow where the streamlines form closed orbits.
We therefore refer to them as particle fixed points in the elliptic
region. We also alternatively refer to them as the finite $\mathbf{r}$
fixed points. Now that we have the exact solution for the fixed points,
we may assess how good the $\mathcal{O}\left(St\right)$ approximation
is, by substituting $\mathbf{v}$ from eq. (\ref{eq: inertial equation})
into (\ref{eq:fixed points exact}). We obtain fixed points, also
shown in figure \ref{fig:Fixed-point-comparison}, that are accurate
to $\mathcal{O}\left(St\right)$, as expected.

\section{Point vortex simulations\label{sec:Point-vortex-simulations}}

\begin{figure}[!h]
\includegraphics[clip,width=0.4\linewidth]{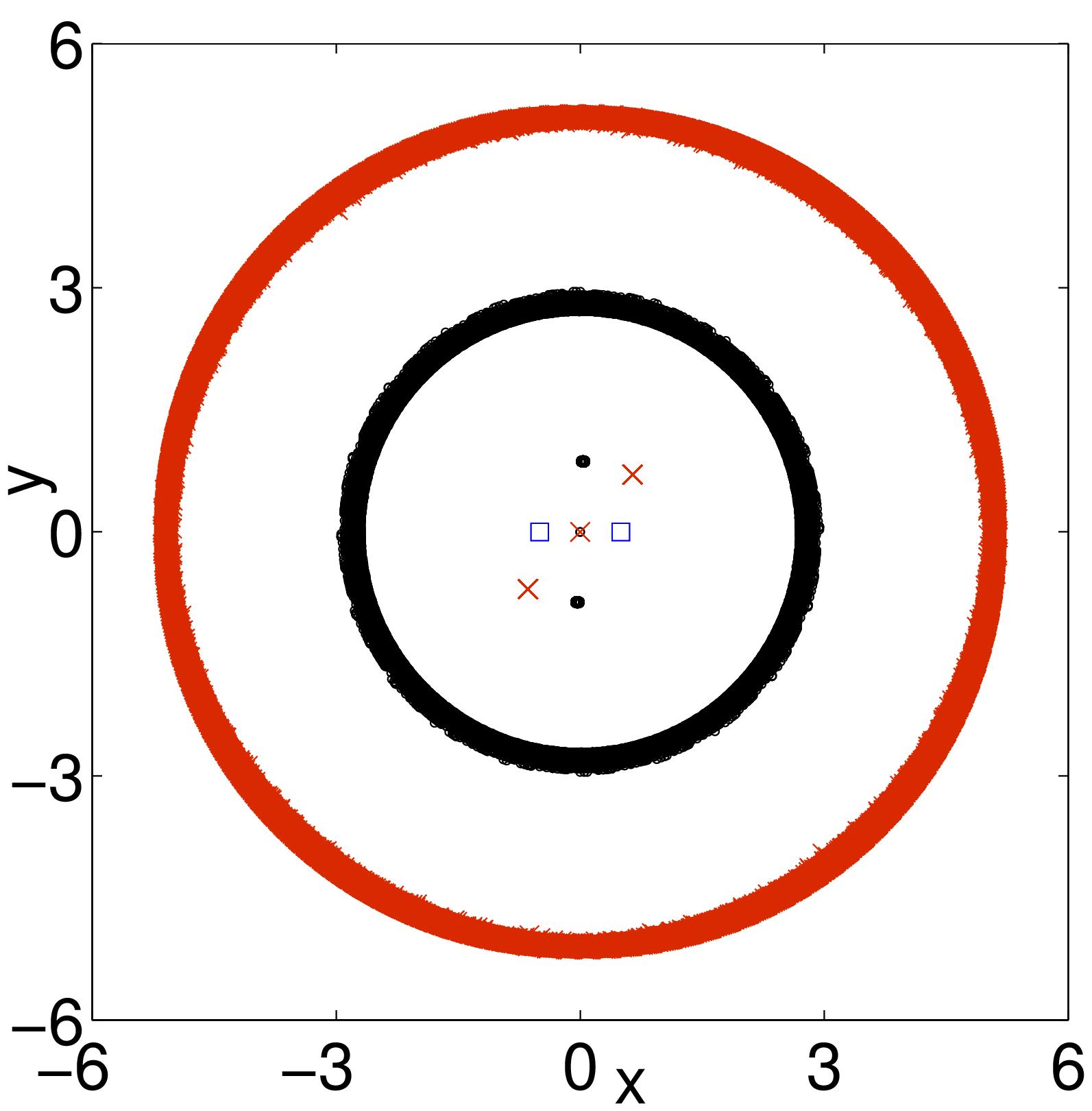} \includegraphics[width=0.5\linewidth]{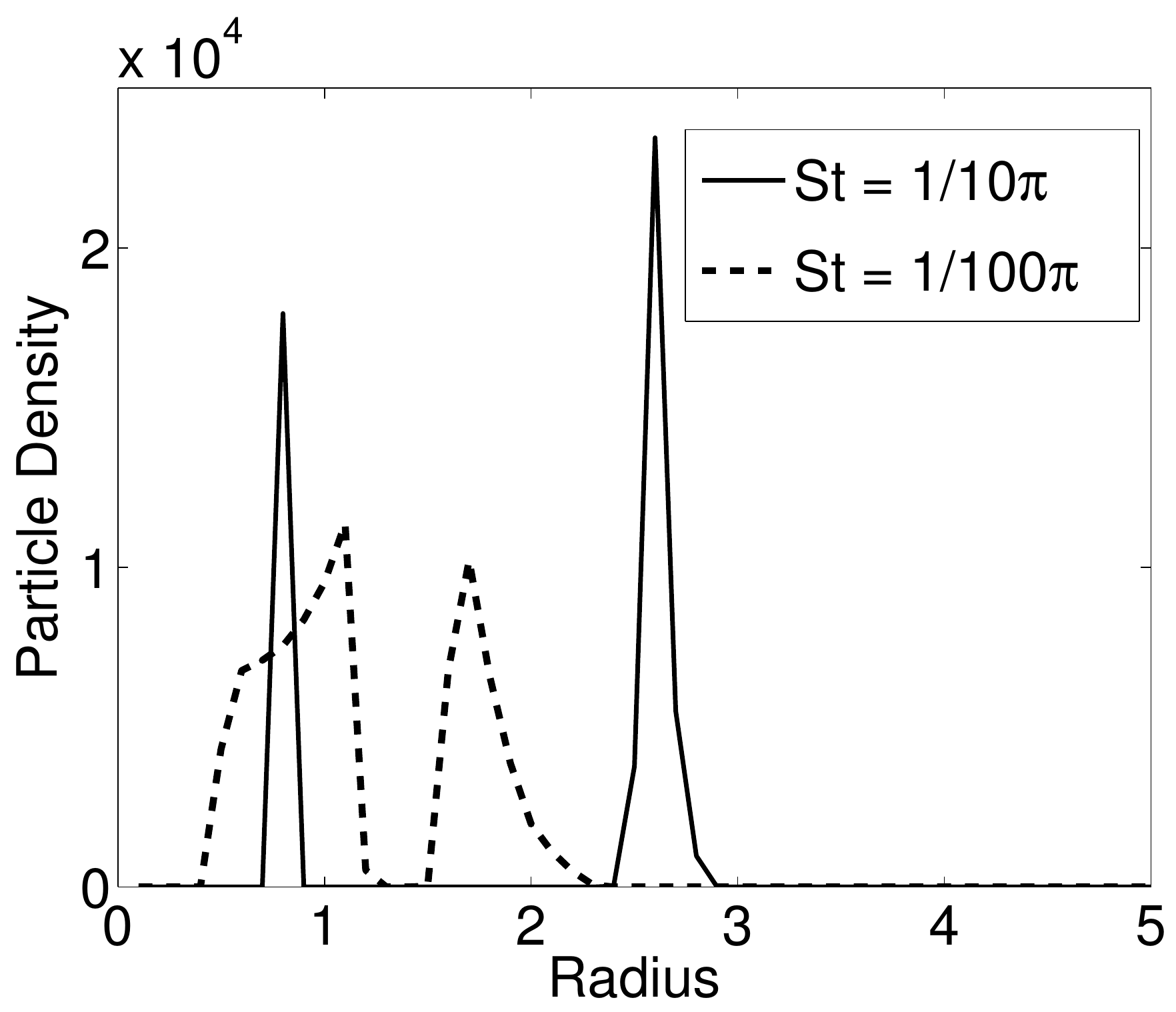}
\caption{\label{fig:Poincare-sections} (Left) Particle locations after $100T$
for $St=1/100\pi$ (black) and $1/7.5\pi$ (red). The fixed points
are the filled circle and the crosses respectively. The phase chosen
is when the point vortices, indicated by blue squares, lie on the
horizontal axis. (In the figure, the thickness of outer ring-clusters
is exaggerated.) (Right) Particle density versus radius after 10 time
periods. The density profiles narrow with time. At a higher Stokes
number, the particles are more sharply clustered. Also as Stokes increases,
the nonlinear wave propagates outwards faster, so the second peak
is seen at a bigger radius.}
\end{figure}

We perform inviscid numerical simulations for a range of Stokes numbers.
We start with $5\times10^{4}$ particles uniformly distributed over
a region encompassing the invariant manifolds. We use $\Gamma=2\pi$
unless otherwise mentioned. Figure \ref{fig:Poincare-sections} shows
the particle locations after $100$ time periods of rotation, for
two Stokes numbers. In each case there are three fixed points at which
heavy particles cluster: the fixed point at $|\mathbf{r}|=\infty$
(which exists because any nonzero inertia in a heavy particle centrifuges
the particle out from a region of rotating motion), as well as the
symmetric pair we have discussed. This pair, lying within the elliptic
flow region, is indicated in the figure, and although not apparent,
a very large number of particles are clustered at these. The new fixed
points for inertial particles are thus attracting fixed points. The
other fixed point, $|\mathbf{r}|=\infty$, attracts all particles
which begin at a large distance away from the origin, since for them
the system may effectively be replaced by a single vortex of twice
the circulation, i.e., $2\Gamma$. The evolution of these particles
would take the form of a nonlinear wave \cite{Raju1997}, with a clumping
that depends on Stokes number. In addition, the $|\mathbf{r}|=\infty$
fixed point has a basin of attraction within our region of interest.
Particles which will travel towards $|\mathbf{r}|=\infty$ are seen
as a circle of clustered points in figure \ref{fig:Poincare-sections}(left).
The radius of this cluster is a slowly increasing function of time,
going as $r\sim t^{1/4}St$ at large $r$. This is because, at large
$r$, the two vortices act predominantly act as one larger vortex
of twice the strength and the equation of motion in radial coordinates
for particles around a single vortex is $dr/dt\sim St/r^{3}$ (see,
e.g., \cite{Raju1997}). That particles form dense clusters at the
fixed points is seen in the radial density profile of figure \ref{fig:Poincare-sections}(right),
at a particular instant of time. Two regions of clumping are evident.
The inner one corresponding to the elliptic-region fixed points remains
there, but the outer spike slowly moves towards $|\mathbf{r}|=\infty$.
With time, the clumps become narrower and taller. With increasing
inertia too, the clumps become sharper, but not necessarily taller.
At higher inertia, the profile increasingly resembles a double spike.
The total number of particles clustered at the inner fixed points
decreases with inertia, as we shall discuss soon, and at some Stokes
number we no longer have any fixed points at finite $r$.

Applying the inertial equation (eq.\ref{eq: inertial equation}) to
this problem, rather than the complete force balance, has the effect
of omitting the second time derivative of the radial location of a
particle, and for a single vortex Raju \& Meiburg \cite{Raju1997}
show that this approximation becomes exact at long times.

\begin{figure}
\includegraphics[width=6.4cm]{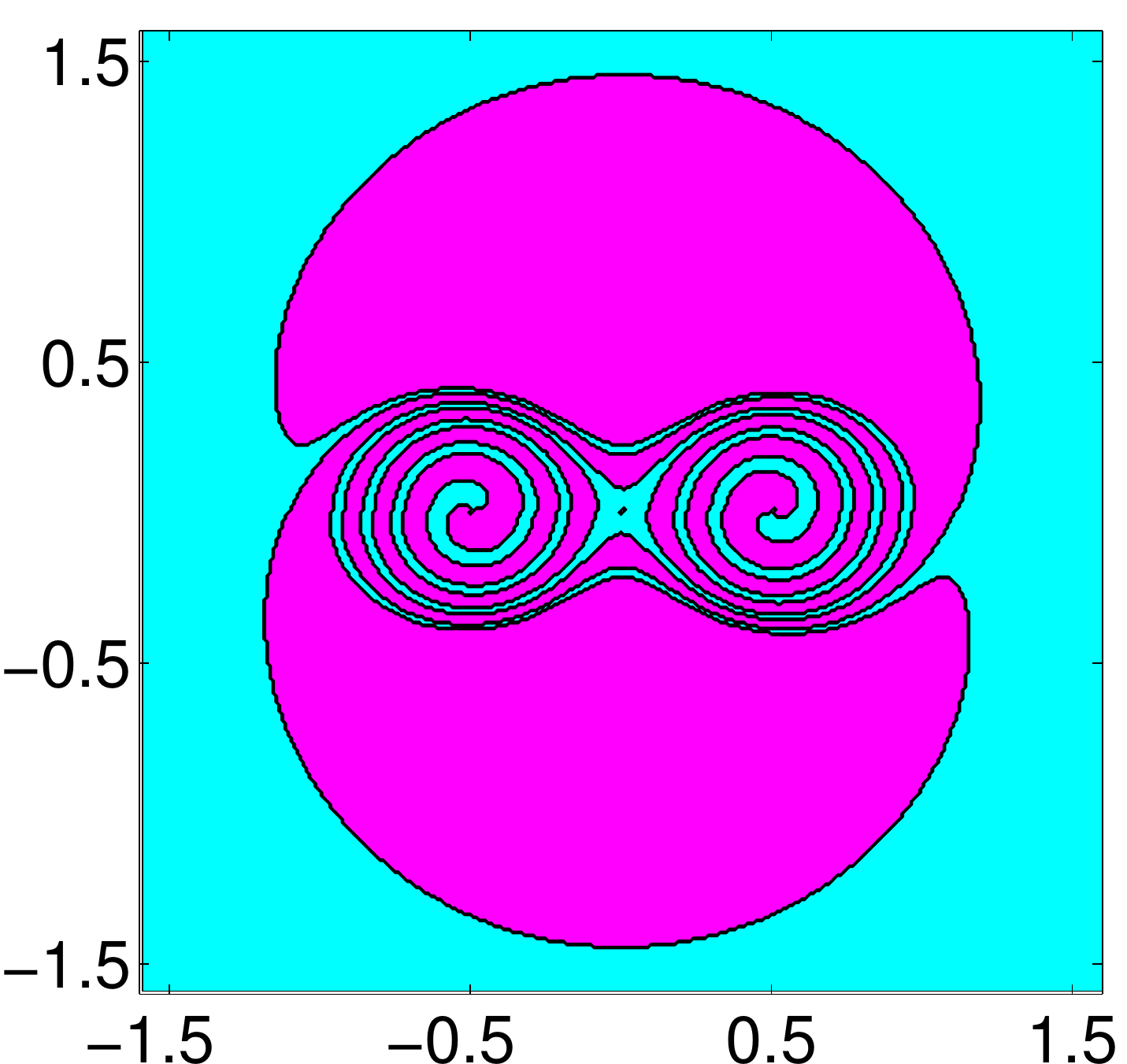}
\includegraphics[width=6.4cm]{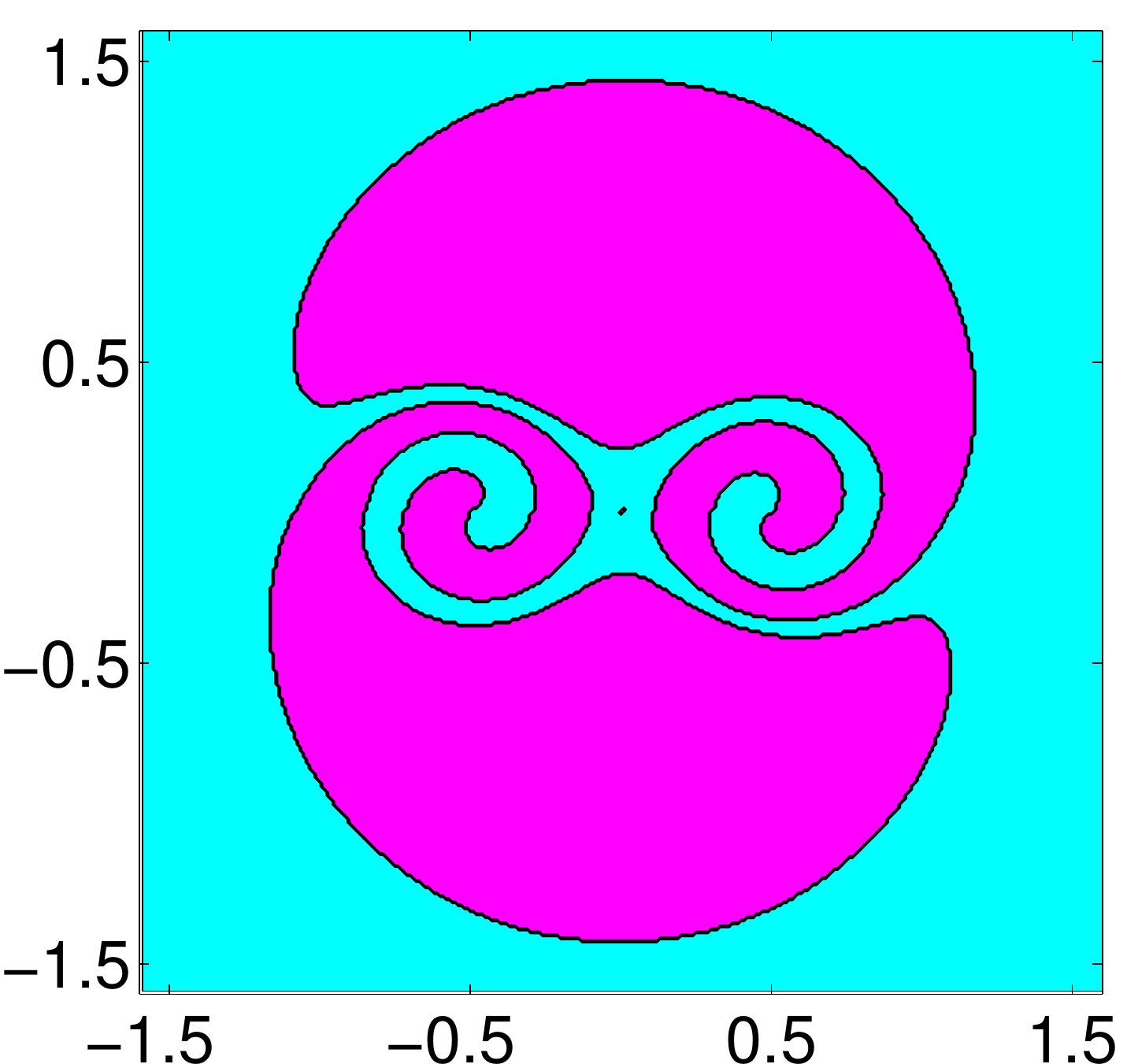}
\includegraphics[width=6.4cm]{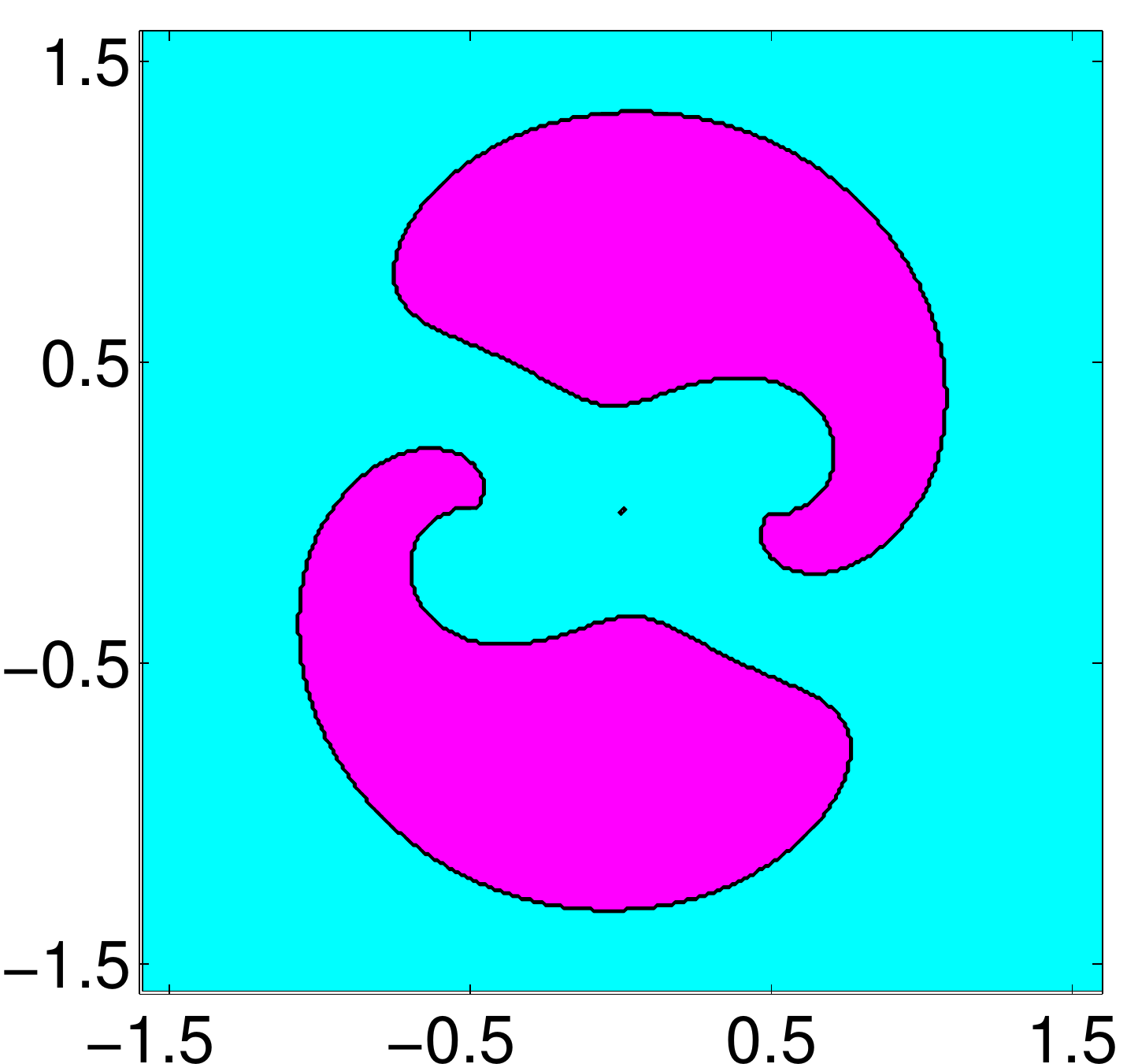}
\includegraphics[width=6.4cm]{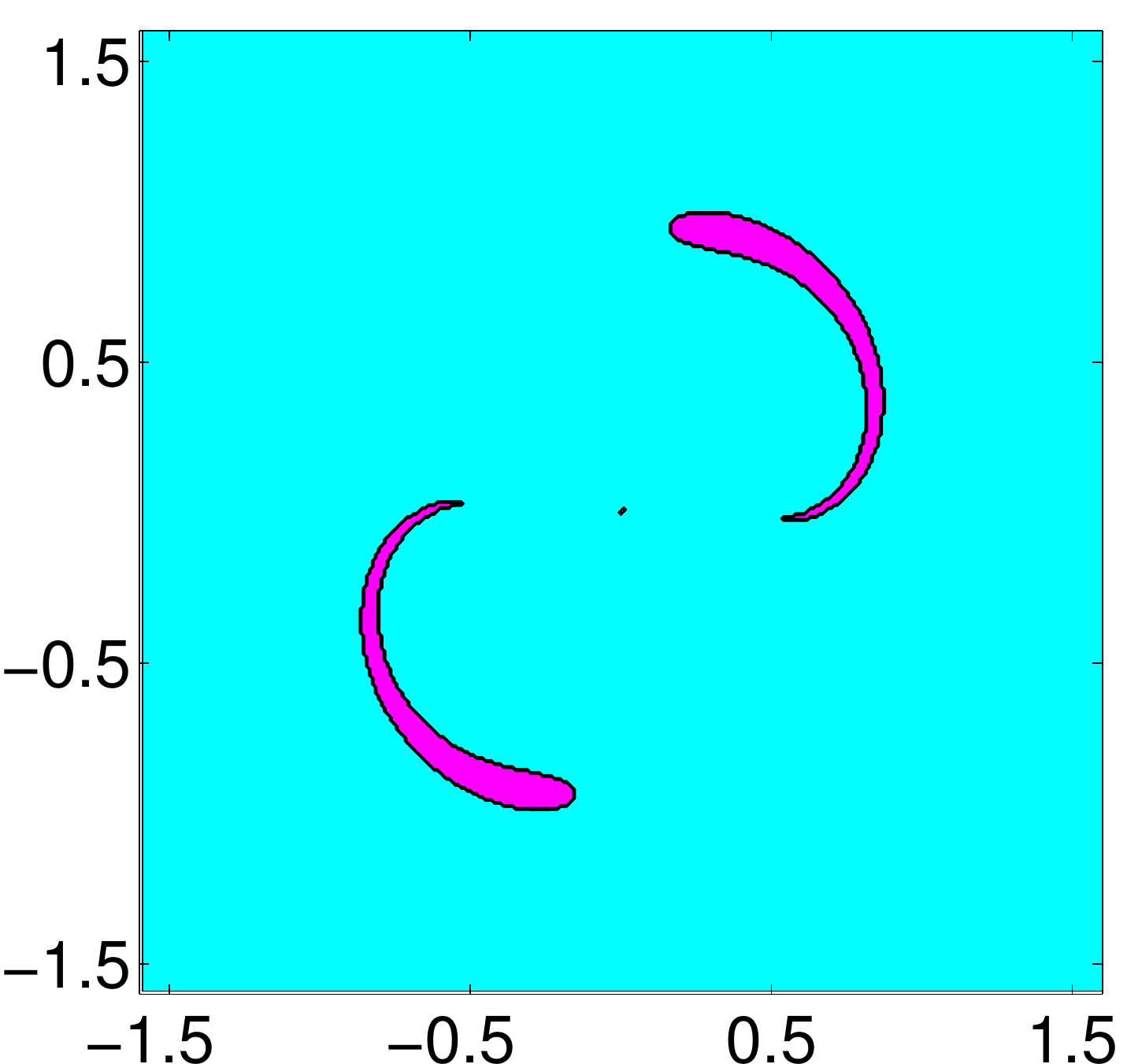}
\caption{Basins of attraction of the fixed points for $St=1/300\pi$ (top left),
$1/100\pi$ (top right), $1/20\pi$ (bottom left), and $1/7.5\pi$
(bottom right). The red region shows the basin of attraction of one
of the finite $\mathbf{r}$ fixed points. Particles in the blue region
escape to infinity. \label{fig:Basin-boundaries}}
\end{figure}

Basin boundaries of each fixed point are shown in figure \ref{fig:Basin-boundaries}.
Particles which begin their lives inside the regions shown in red
are attracted to one of the fixed points at finite radius and particles
in the other regions drift away to infinity. As the Stokes number
increases, the basin of attraction of the finite $r$ fixed points
shrinks in size. These basins of attraction disappear completely for
$St>St_{cr}$, along with the fixed points themselves. At very low
inertia, the spirals are very tightly wound. It may be argued from
the radial velocity of an inertial particle moving around a single
vortex that the spacing between two crossings must be proportional
to $St$. Thus two particles which begin life very close to each other,
but in the basin boundaries of different attractors, have vastly different
fortunes. One gets trapped forever in the vicinity of the vortices
while the other is slowly centrifuged out to infinity.

\begin{figure}
\noindent \includegraphics[width=9cm]{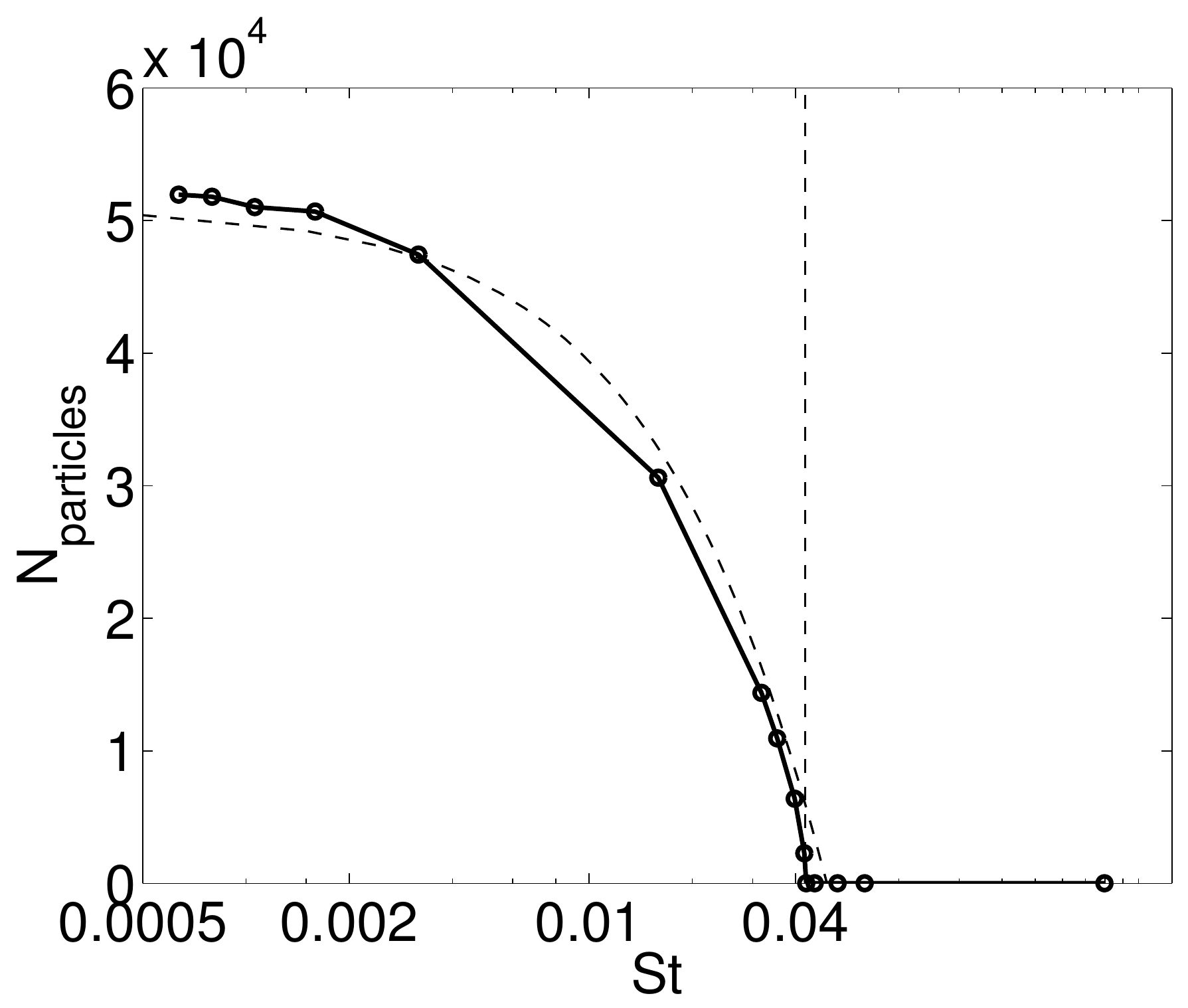} \caption{The number of particles in each simulation that are attracted to the
finite $\mathbf{r}$ fixed points. This number is proportional to
the area of the basin of attraction. A quadratic fit is provided just
to guide the eye. The vertical line indicates $St=St_{cr}$.\label{fig:n_inside}}
\end{figure}

In figure \ref{fig:n_inside}, we plot the number of particles starting
from a uniform grid over a domain $\left(-1.6,1.6\right)\times\left(-1.6,1.6\right)$
that are attracted to one of the two fixed points. Because we started
with uniformly distributed points, this is proportional to the area
of its basin of attraction. It is seen that close to the critical
Stokes number, the rate of decrease of the number of particles is
very high, the graph being practically vertical, indicative of a finite-Stokes
singularity. The reason for this singular behavior is not clear to
us at this point. 

\subsection{Linear analysis near the particle fixed points\label{sub:Linear-analysis}}

Behavior in the vicinity of the inertial particle fixed points is
obtained by linearising the particle dynamics in the rotating frame
(eq. \eqref{eq:Particle_motion_rotating_frame}) along the lines of
\cite{Bec2005}. At the fixed point, we have $\hat{\mathbf{v}}_{fp}=0$
and $\hat{\mathbf{u}}_{fp}=-St\mbox{\ensuremath{\left(\Omega^{2}\hat{\mathbf{r}}\right)}}$.
This gives, for the linear perturbations,

\begin{align}
\frac{d\left(\delta\hat{\mathbf{x}}\right)}{d\hat{t}} & =\delta\hat{\mathbf{v}}\nonumber \\
\frac{d\left(\delta\hat{\mathbf{v}}\right)}{d\hat{t}} & =\frac{\delta\hat{\mathbf{u}}-\delta\hat{\mathbf{v}}}{St}-2\bm{\Omega}\times\left(\delta\mathbf{\hat{v}}\right)+\Omega^{2}\delta\mathbf{\hat{r}}\label{eq:Linearised_particle_dynamics}
\end{align}

\begin{figure}[H]
\noindent \includegraphics[width=8cm]{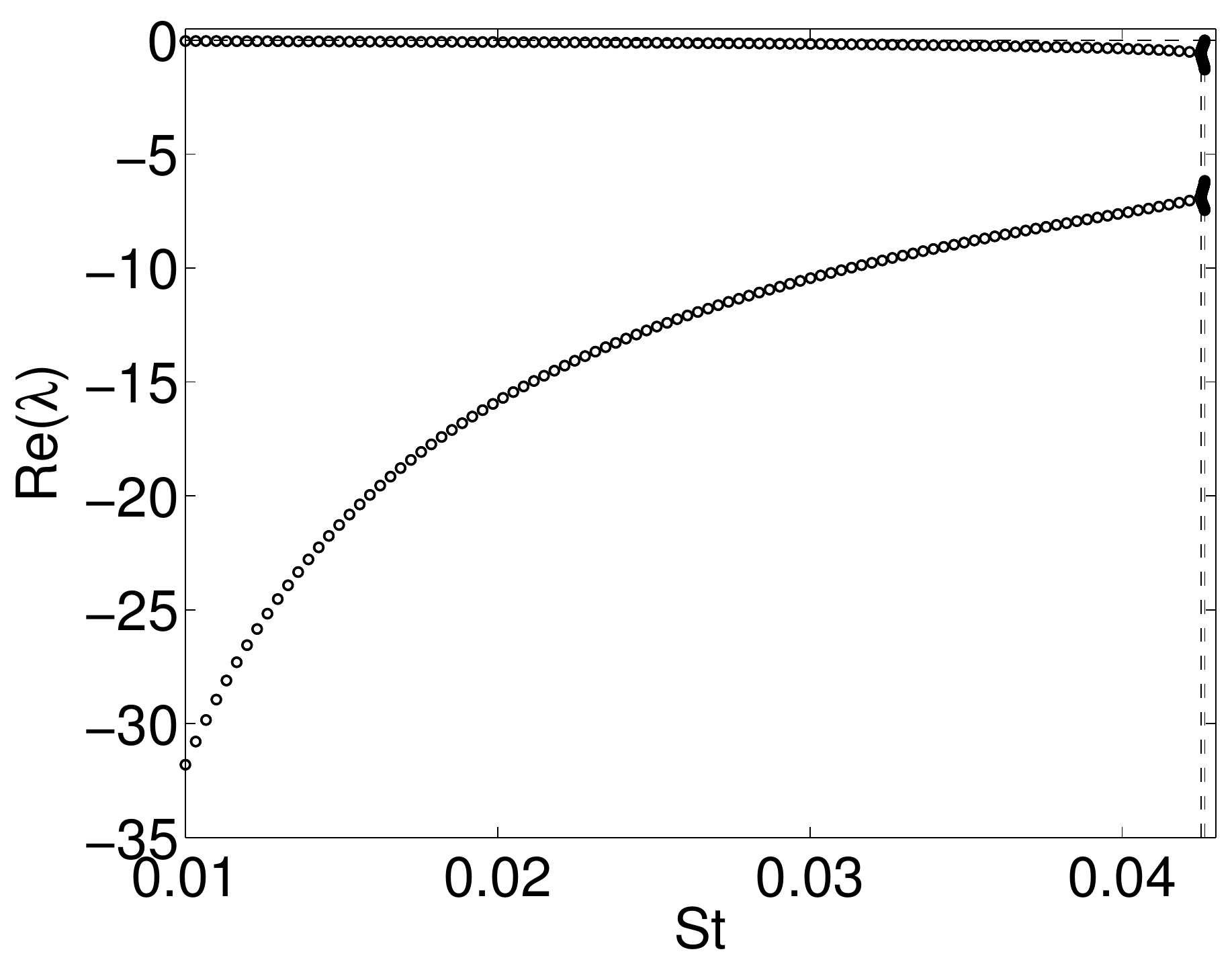} \includegraphics[width=8cm]{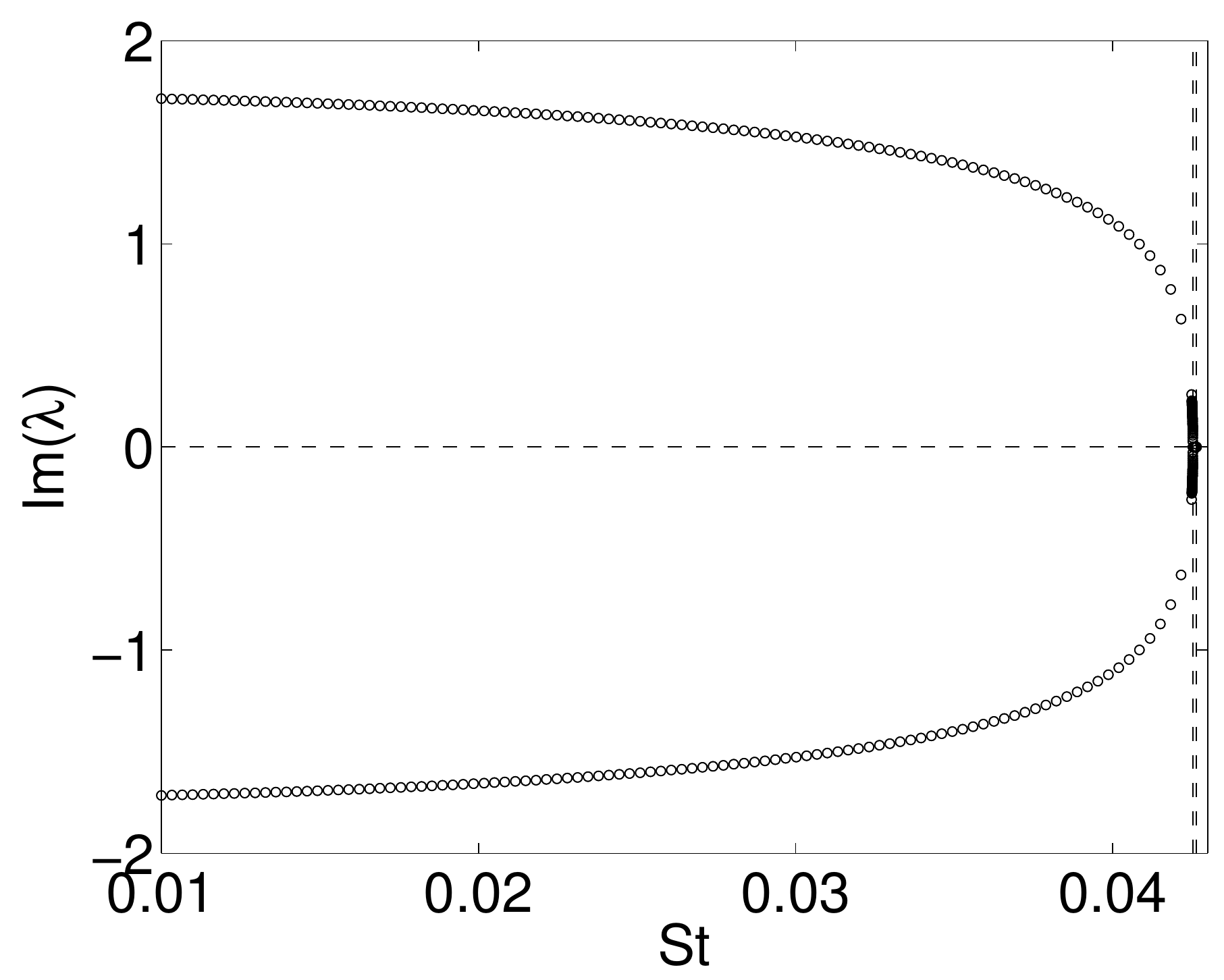}

\noindent \includegraphics[width=8cm]{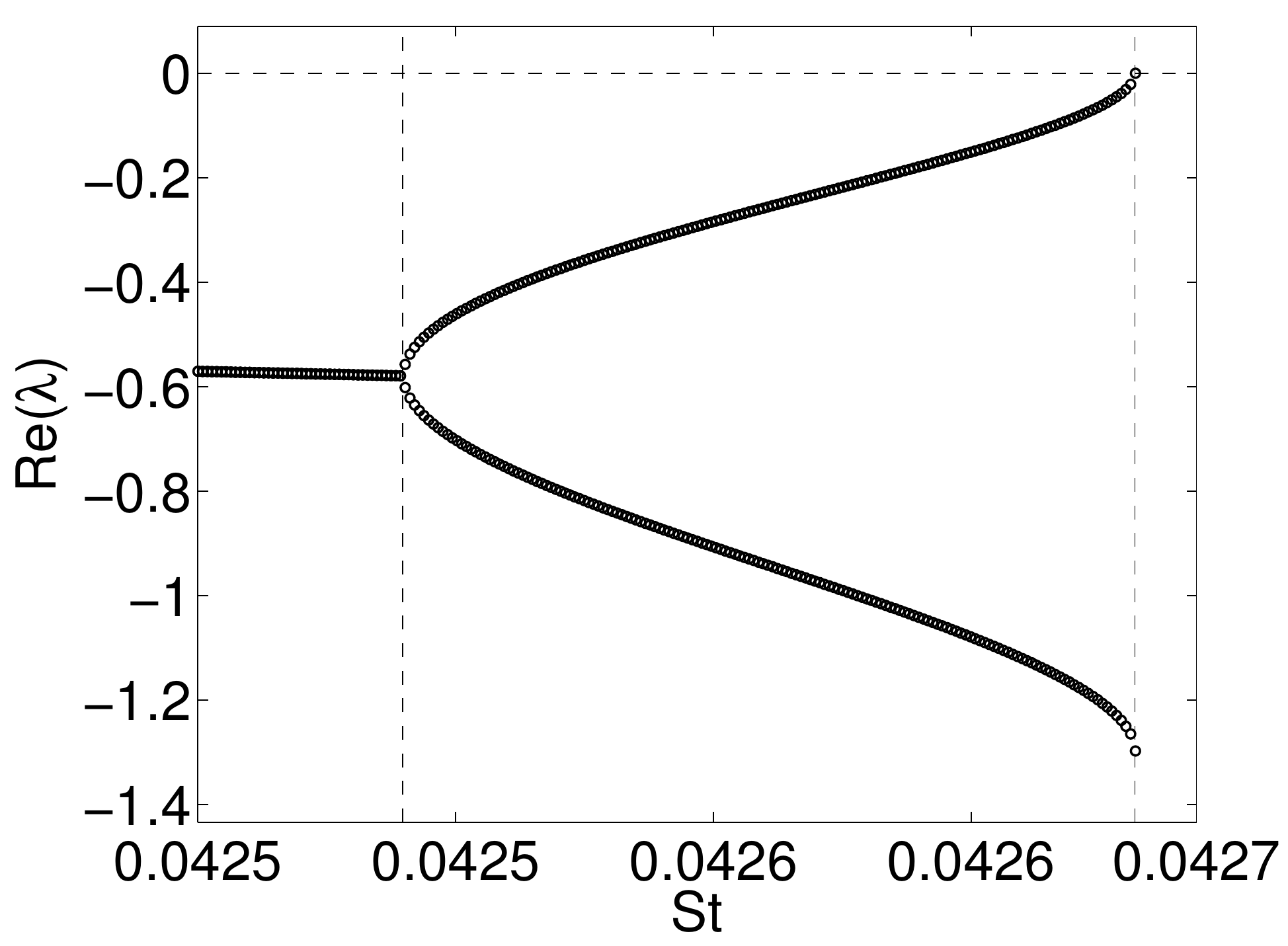}\includegraphics[width=8cm]{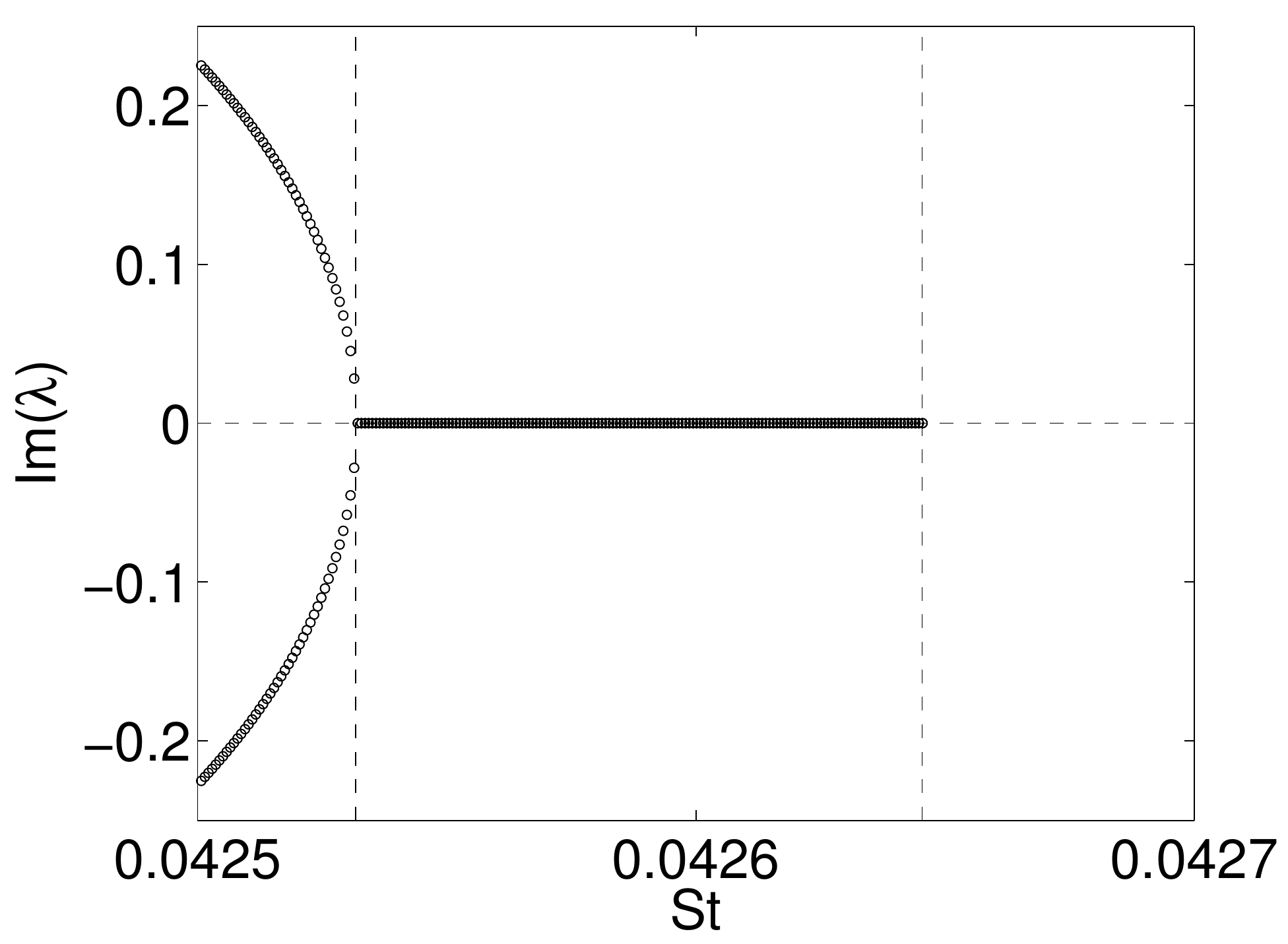}

\noindent \caption{Real and imaginary parts of the eigenvalues of eq. \ref{eq:Linearised_particle_dynamics}.
The two vertical dashed lines are at values of $St$ where a) the
two pairs complex conjugate eigenvalues become wholly real, and b)
at $St=St_{cr}$. The values for these Stokes numbers may be found
in the text. The figures at the bottom are zoomed-in near $St_{cr}$.
The real part crosses zero (at an infinite slope) at $St_{cr}$. \label{fig:fixed-point-bifurcation}}
\end{figure}

The derivatives on the LHS of eq. \eqref{eq:Linearised_particle_dynamics}
are derivatives along particle trajectories. The perturbation fluid
velocities in the RHS are calculated by differentiating the Biot-Savart
law. The eigenvalues of the linearised velocity equations are plotted
in figure (\ref{fig:fixed-point-bifurcation}). One of the eigenvalues
crosses zero at the crititcal Stokes number. It is also worth noting
that the slopes of the eigenvalue curves diverge at $St=St_{cr}.$

\begin{figure}
\noindent \includegraphics[width=8cm]{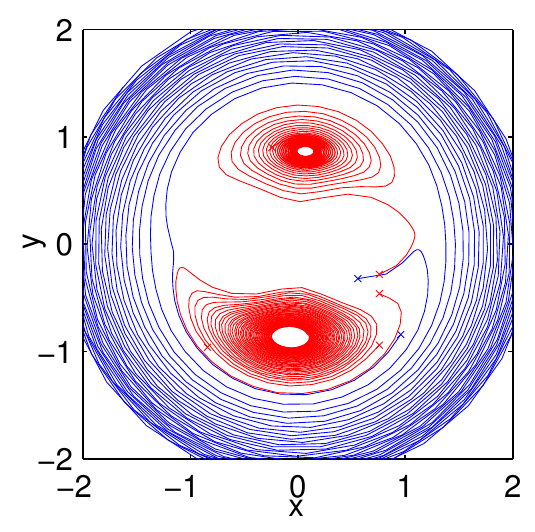} \includegraphics[width=8cm]{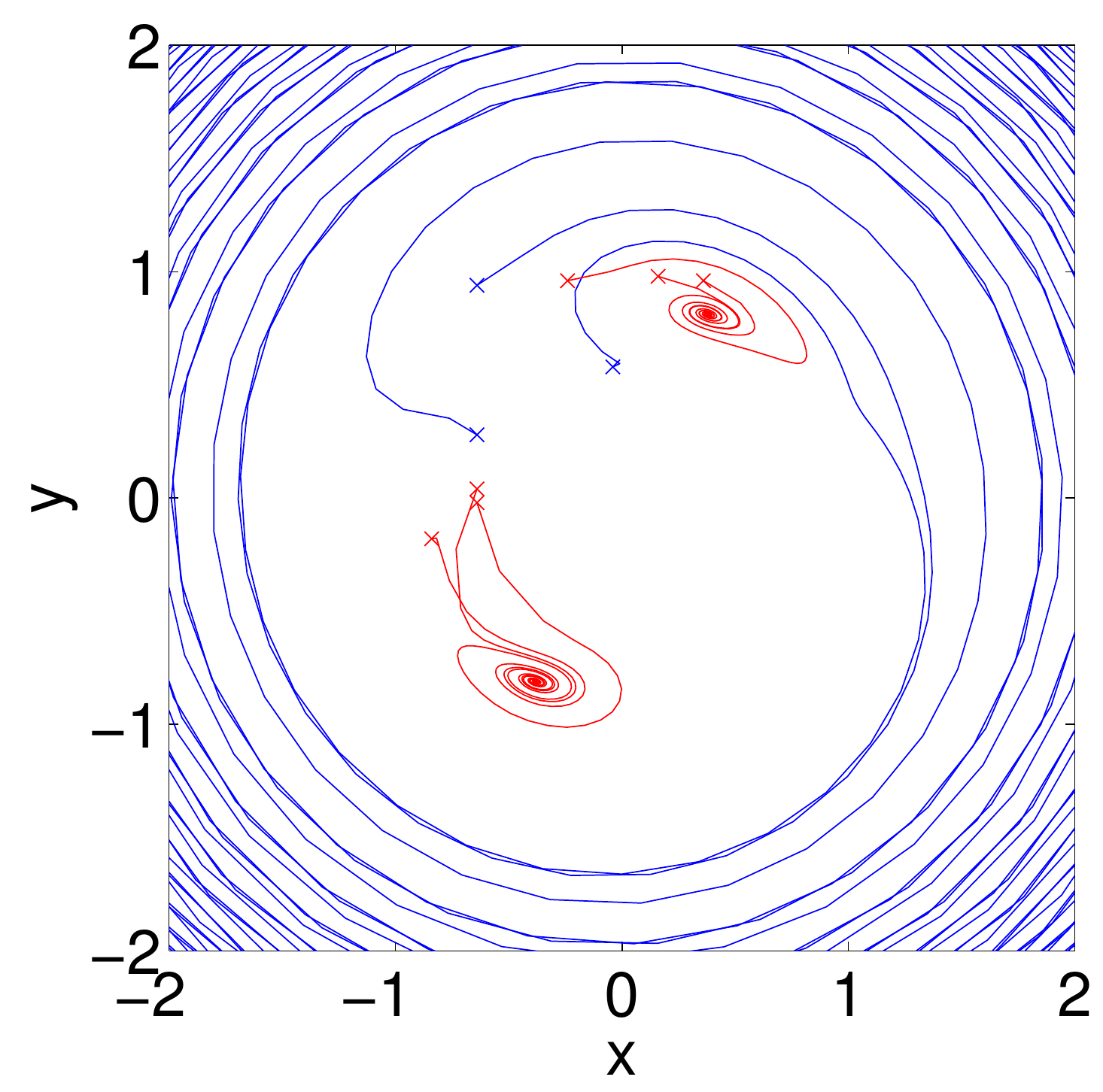}

\caption{\label{fig:Spiral Nodes}Particle trajectories for representative
Stokes numbers of $St=1/50\pi$ (left) and $St=1/10\pi$ (right) showing
that the fixed points in the rotating frame are spiral nodes. The
crosses indicate starting positions for the particles. Trajectories
in blue leave the vicinity of the vortices and centrifuge out to infinity.
Trajectories in red fall into one of the fixed points.}

\end{figure}

At $\mathcal{O}\left(St\right)$, a negative Okubo-Weiss parameter
is a necessary condition for the existence of an attracting fixed
point. We have seen from the argument of Sapsis \& Haller \cite{Sapsis2010}
that this cannot happen in a neighborhood of closed streamlines in
a fixed frame of reference. Things are different in a rotating frame:
we may have a negative divergence of particle velocity in a neighborhood
of streamlines that are closed paths in the rotating frame. The velocity
in the rotating frame is (quantities with $\hat{}$s are measured
in the rotating frame) 
\[
\hat{\mathbf{v}}=\hat{\mathbf{u}}-\tau\left(\frac{d\hat{\mathbf{u}}}{d\hat{t}}+2\vec{\Omega}\times\hat{\mathbf{u}}-\Omega^{2}\hat{\mathbf{r}}\right)\mbox{,}
\]

giving, for the divergence of the particle velocity,

\[
\hat{\nabla}\cdot\hat{\mathbf{v}}=-\tau\left(\frac{\partial\hat{u}_{i}}{\partial\hat{x}_{j}}\frac{\partial\hat{u}_{j}}{\partial\hat{x}_{i}}+2\hat{\nabla}\cdot\left(\bm{\Omega}\times\hat{\mathbf{u}}\right)-2\Omega^{2}\right)=-\tau\left(\frac{\partial\hat{u}_{i}}{\partial\hat{x}_{j}}\frac{\partial\hat{u}_{j}}{\partial\hat{x}_{i}}-2\bm{\Omega}\cdot\hat{\bm{\omega}}-2\Omega^{2}\right)
\]
\[
\hat{\nabla}\cdot\hat{\mathbf{v}}=-\tau\left(\frac{\partial\hat{u}_{i}}{\partial\hat{x}_{j}}\frac{\partial\hat{u}_{j}}{\partial\hat{x}_{i}}+2\Omega^{2}\right)\mbox{, }
\]
where $\hat{\bm{\omega}}=-2\Omega\mathbf{e}_{z}$ is the vorticity
in the rotating frame and the summation convection has been used.
The first product on the right hand side may be shown to be equal
to $S_{ij}S_{ij}-2\Omega^{2}$, making the divergence of particle
velocity equal to 
\begin{equation}
\nabla\cdot\mathbf{\hat{v}}=St{\hat{Q}},
\end{equation}
 where $\hat{Q}$ is the Okubo-Weiss parameter in the rotating frame.

\begin{figure}
\includegraphics[width=9cm]{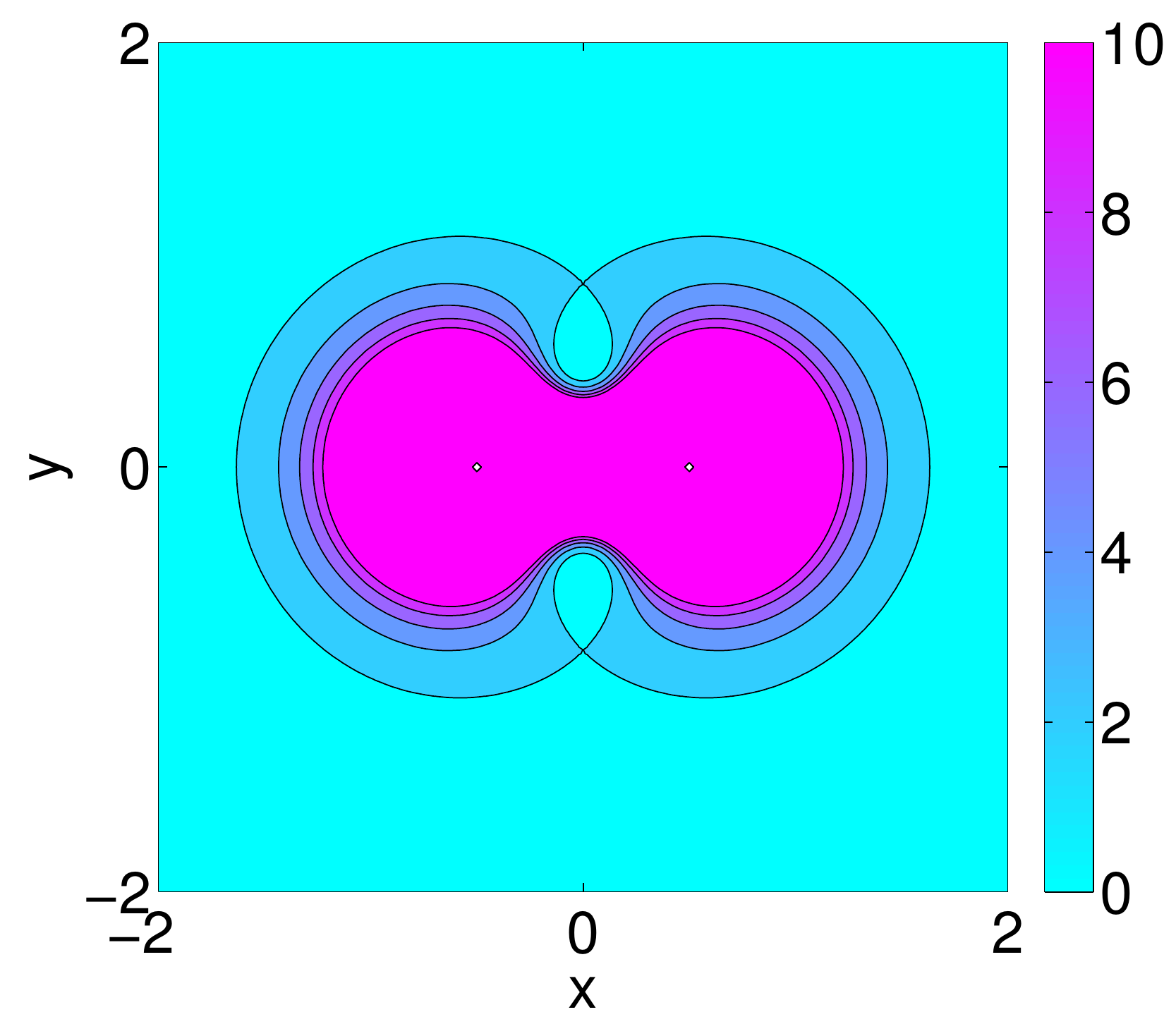} 
\caption{Contour-plot of $S_{ij}S_{ij}$. The negative of this quantity, $\hat{Q}$,
is the Okubo-Weiss parameter, which is proportional to the divergence
of particle velocity. \label{fig:rotational-okubo-weiss}}
\end{figure}

The quantity $(S_{ij}S_{ij})$ is plotted in figure \ref{fig:rotational-okubo-weiss},
and seen to be positive in the vicinity of the attracting fixed point,
making the Okubo-Weiss parameter negative; we thus satisfy the necessary
condition for the existence of attracting fixed points, in a neighborhood
where fluid particles follow elliptical streamlines.

\section{Navier-Stokes Simulations}

We now substantiate our results by studying the above flow in a more
realistic setting: by including viscosity and beginning with Lamb-Oseen
vortices, in contrast to the point vortices of section III. The flow
obeys the two-dimensional Navier-Stokes equations 
\begin{equation}
D_{t}\omega=\nu\nabla^{2}\omega\mbox{ , }-\nabla^{2}\psi=\omega,\label{eq:NS}
\end{equation}
 where $\omega$ is the vorticity, $\nu$ is the kinematic viscosity,
$\psi$ is the streamfunction, i.e. ${\bf u}=(-\partial_{y}\psi,\partial_{x}\psi)$,
and $D_{t}\equiv\partial_{t}+{\bf u}\cdot\nabla$ is the material
derivative. Particles obey eq. (\ref{eq:particle motion}). We use
a square domain with each side of length $L=2\pi$ and employ periodic
boundary conditions. Eqs.\eqref{eq:NS}, and the continuity equation
are numerically integrated using a pseudo-spectral method. Time-advancement
is done using an exponential Adams-Bashforth scheme. Space is discretized
with $N^{2}$ collocation grid points. We have verified (not shown
here) that grid-convergence is achieved by conducting simulations
with varying grid resolutions $N=128$ and $N=256$ and obtaining
the same results. In what follows, we have used $N=256$ unless stated
otherwise.

We initialize the simulation with two Gaussian vortices positioned
at $(x_{1},y_{1})$ and $(x_{2},y_{2})$, of vorticity 
\begin{eqnarray}
\omega_{1}(x,y) & = & \omega_{0}\exp[-(r_{1}/r_{0})^{2}]\ {\rm {and}}\\
\omega_{2}(x,y) & = & \omega_{0}\exp[-(r_{2}/r_{0})^{2}].
\end{eqnarray}
 Here, $r_{0}$ is the width of the vortex, $\omega_{0}$ denotes
the vortex amplitude, $r_{1}^{2}=(x-x_{1})^{2}+(y-y_{1})^{2}$ and
$r_{2}^{2}=(x-x_{2})^{2}+(y-y_{2})^{2}$ and the separation between
the centers of the vortices $\ell\equiv(y_{2}-y_{1})^{2}+(x_{2}-x_{1})^{2}$
is kept fixed at $\ell=0.9818$ (we obtain this value because we choose
an integer number of grid-spacings). We fix $r_{0}=\pi/32$, $\omega_{0}=2^{11}/\pi\approx652$
(these values are tuned so that the timeperiod of rotation is $T\approx1$),
$\nu=10^{-4}$ giving $Re=10^{4}$, and vary $St$ from $0$ to $0.05$.
We start the simulation with $N_{p}=10^{3}$ particles distributed
randomly over the entire domain and study how particle distribution
evolves.

\begin{figure}[!h]
\includegraphics[width=9cm]{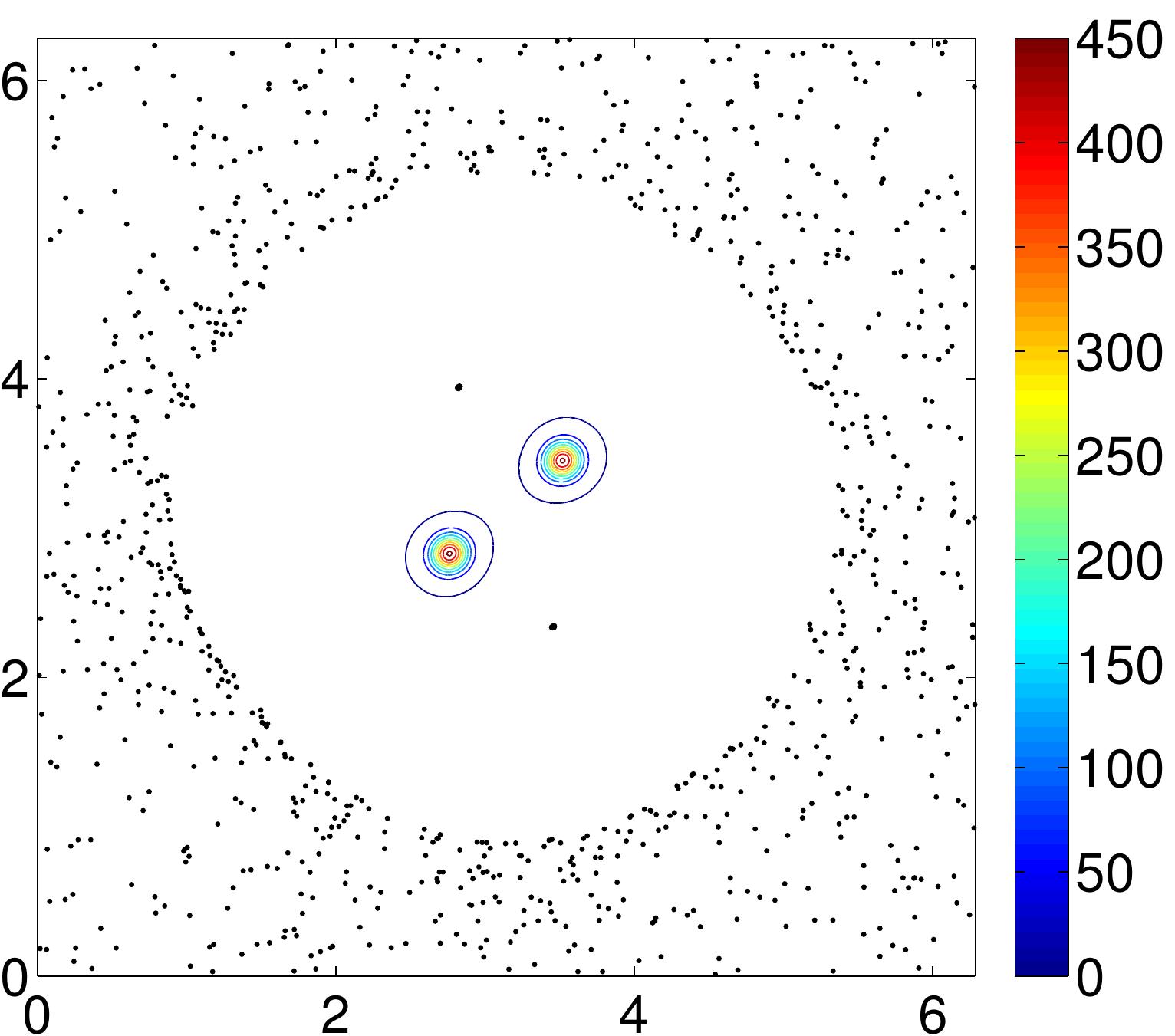}\includegraphics[width=9cm]{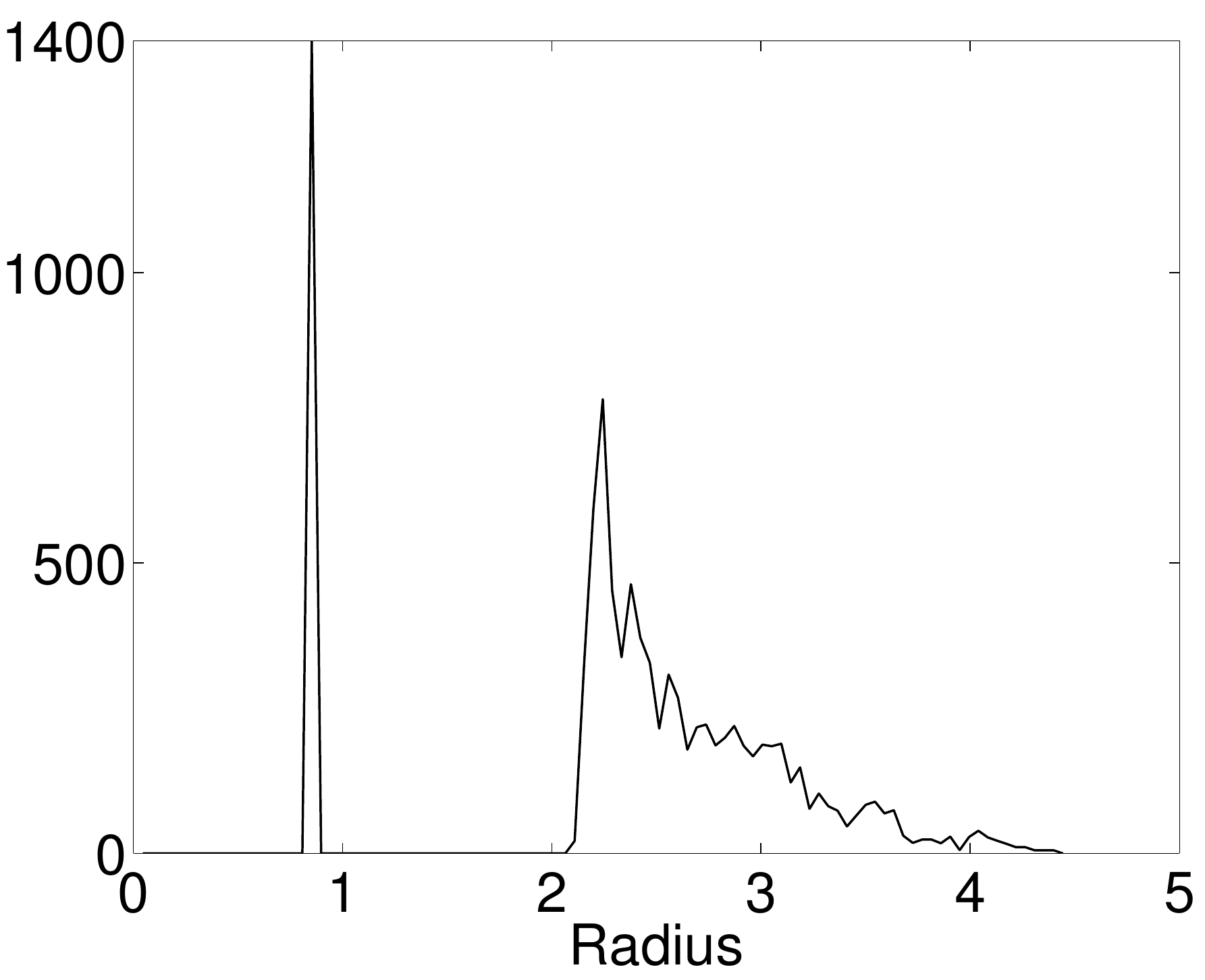}
\caption{LEFT: Plot of particle positions and vorticity contours at $t=10$
for $St=0.027$. The fixed point can be seen at ($\pi\pm0.32$,$\pi\pm0.8)$.\protect \\
RIGHT: Plot of particle density showing versus radius.\label{fig:Particles_vorticity_t10}}
\end{figure}

\begin{figure}[!h]
\includegraphics[width=9cm]{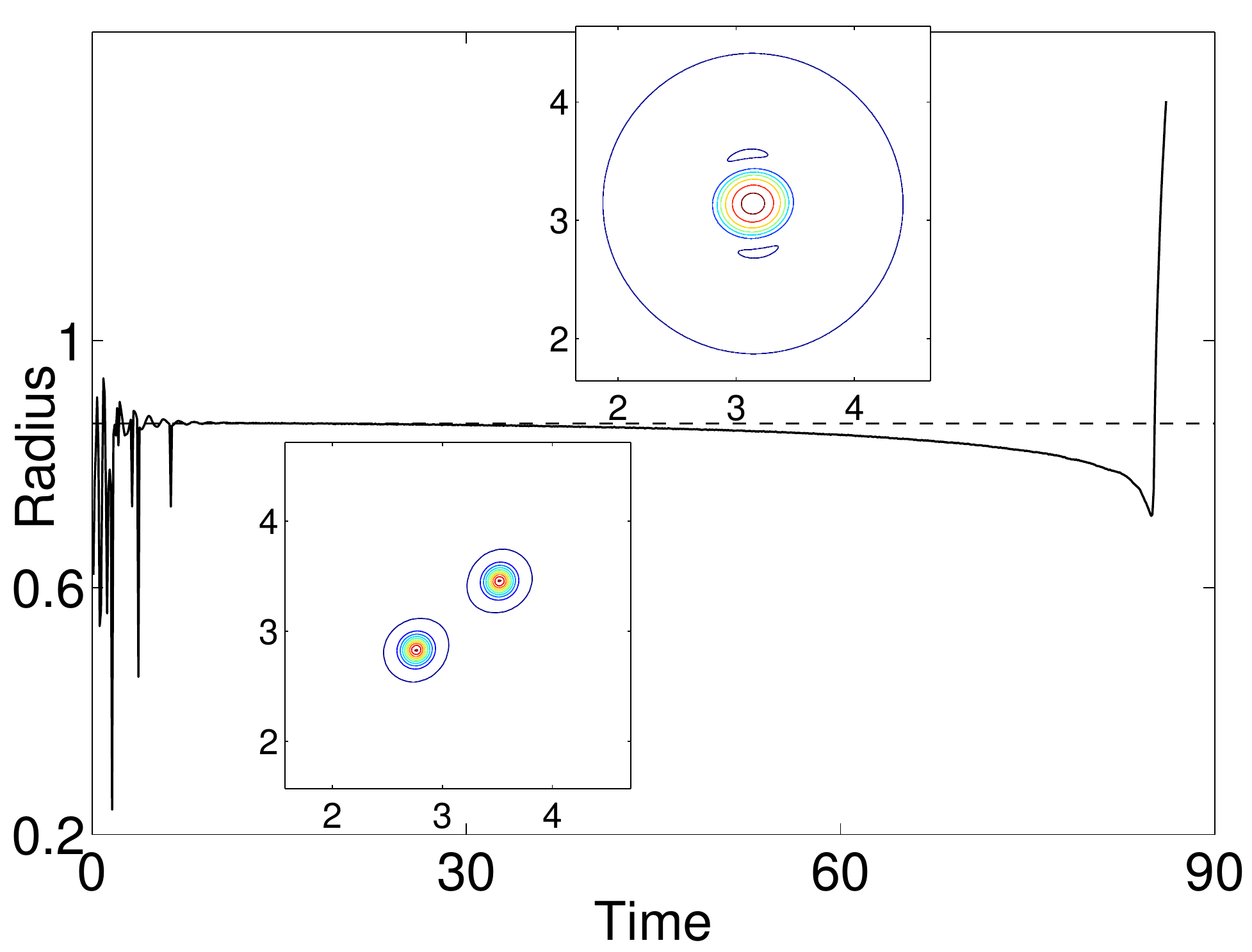} \caption{The radial location of the fixed point time for a viscous simulation.
Inset in the figure are the contours of vorticity at $t=10$ and $t=86$
for $St=0.03$. The maxima of vorticity are at $\omega=452$ and $\omega=132$
respectively. The dashed line is $r=\sqrt{3}/2$ \label{fig:radial_location_viscous_fixed_point}.
Since we are interested to look at times equal to and larger than
merger time, we use $N^{2}=1024^{2}$ collocation points.}
\end{figure}

\begin{figure}[!h]
\includegraphics[width=9cm]{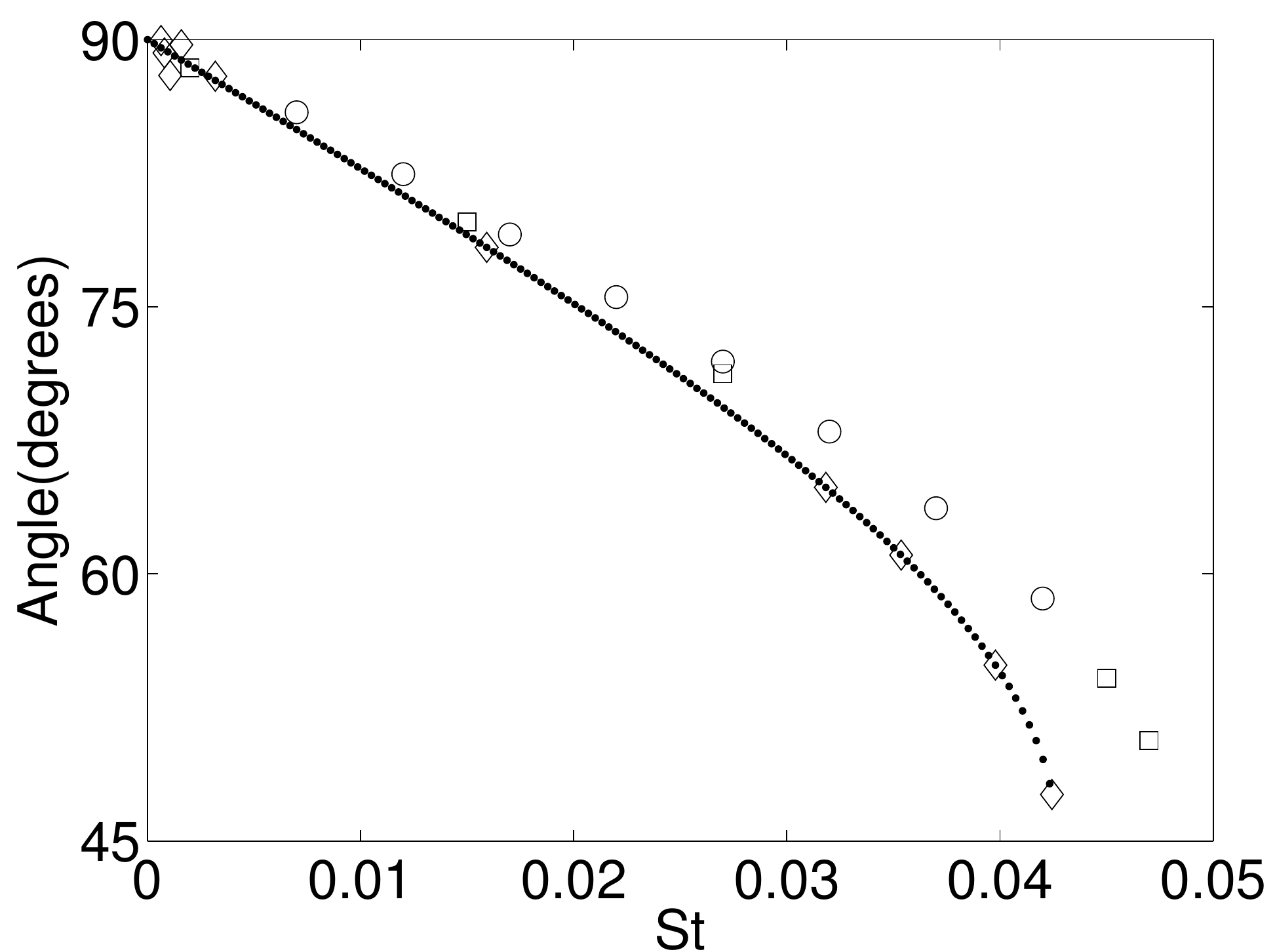} \caption{Fixed point predictions (at $t=50$). We plot the angle the line joining
the fixed points makes with the line joining the centres of the vortices.
The dots are exact solutions. The diamonds are point-vortex solutions.
The circles and squares are DNS results for grid resolutions of $256^{2}$
and $512^{2}$ respectively. \label{fig:Comparison_viscous_inviscid}}
\end{figure}

An identical same-signed vortex pair will ultimately merge, In the
first stage of this process, the vortices go around each other in
a fashion very similar to the inviscid case. During this phase, they
also diffuse out slowly. When the radius of the vortices reaches a
quarter of the separation between them, a second stage in the merger
process begins, with a significant and sudden increase in the radial
component of the velocity of the vortices, drawing them towards each
other. The Reynolds number in this simulation is $Re=10^{4}$, which
means, given the initial radius, that the vortices complete about
90 complete cycles, up to a time of $t=86$ before the second stage
of the merger process begins. We note that particles with $St\sim\mathcal{O}\left(1/T\right)$,
(where $T=86$ timeperiods) would have fallen into the fixed points
before the merger process begins. The cluster at the fixed point thus
has a long life before the merger process is completed, after which
it begins a slow drift away from the merged vortex. We are interested
here in what happens to the cluster during the first phase of the
merger process. Figure \ref{fig:Particles_vorticity_t10} shows a
snapshot of the particle positions and contours of the vorticity at
$t=10$ for $St=0.027$. The cluster at the fixed point is remarkably
similar to the one seen in the inviscid case, as is evident when one
compares this figure with fig. \ref{fig:Poincare-sections}. Figure
\ref{fig:radial_location_viscous_fixed_point} shows the radial distance
at which particles cluster (measured from the point halfway between
the vortex centres) as a function of time, showing that, for this
Stokes number$=0.03$, (which is below $St_{cr},$) the cluster at
the fixed points is remarkably stable for a long time. Figure \ref{fig:Comparison_viscous_inviscid}
compares the fixed points obtained from the viscous and inviscid simulations.
The critical Stokes number is slightly higher in viscous simulations. 

\section{Conclusion\label{sec:conclusion}}

We have shown that apart from fixed points in the lab frame, clustering
can be governed by fixed points in a rotating frame. Fixed points
for heavy particles of small Stokes number do not coincide with those
for tracer particles, but lie in the vicinity. The location of these
fixed points is within a region where fluid particles follow elliptical
trajectories. Contrary to our understanding of elliptic fixed points
in the lab-fixed frame, these fixed points in a rotating frame can
be attractive, which is the reason for clustering. Note that these
moving fixed points are not limit cycles, because the phase is fixed.

The study here is on a simple model flow, but has relevance to particle
dynamics in turbulence. Persistence times for particles near vortices
are important in turbulent flows. With attracting fixed points in
the vicinity of point vortices, the persistence times are in principle
infinite. In cloud dynamics, for instance, we believe these fixed
points in rotating frames could contribute to droplet clustering and
therefore change the droplet size distributions. Our study indicates
that regions of particle clustering may have to be calculated for
frames of reference that are themselves not fixed. These frame-fixed
points may be expected to have the same effect on the carrier flow
as fixed points in the lab frame. 
\begin{acknowledgments}
The authors wish to thank the two anonymous referees who made very
useful suggestions. The authors wish to thank Jeremie Bec for the
suggestion that led to the stability analysis in section IIIA. 
\end{acknowledgments}
\bibliographystyle{apsrev-title}
\bibliography{2vor_paper}

\end{document}